\documentclass[12pt]{iopart}
\begin{document}

\title[Spin-compatible Quantum gravity from minimum microscopic information]{Spin-compatible construction of a consistent quantum gravity model from minimum information}

\author{P. A. Mandrin}

\address{Department of Physics, University of Zurich, Winterthurerstrasse 190, 8057 
Z\"urich, CH}
\ead{pierre.mandrin@uzh.ch}

\begin{abstract}
This article shows in detail the computations made for the poster presented at the Symposium ''Frontiers of Fundamental Physics'' in July 2014. As was shown in a previous publication, a quantum gravity formulation exists on the basis of abstract quantum number conservation, the laws of thermodynamics, unspecific interactions, and locally maximising the ratio of resulting degrees of freedom per imposed degree of freedom of the theory. The first law of thermodynamics was evaluated by imposing boundary conditions on small volumes of optimised dimension (3+1). As a consequence, no explicit microscopic quantum structure was required in order to recover all well established physics as special cases (Quantum Field Theory, QFT, and General Relativity, GR) and compute all measurable quantities. This article presents the generalised action in terms of tetrads and shows how this action may be related to the spin of generalised matter fields, especially for fermionic matter.
\end{abstract}

%\pacs{04.60.-m, 05.30.-d, 04.20.Cv, 03.70.+k, 45.20.Jj}
\noindent{\it Keywords}: Quantum Gravity, Entropy, General Relativity, Quantum Mechanics
%\submitto{\CQG}
%\maketitle

%%%%%%%%% 1. Introduction

\section{Introduction}
\label{intro}

In \cite{Mandrin1}, a quantum gravity formulation has been shown to exists on the basis of abstract quantum number conservation, the laws of thermodynamics, unspecific interactions, and locally maximising the ratio of resulting degrees of freedom per imposed degree of freedom of the theory. This approach will be referd to as Minimum Information Quantum Gravity (MIQG). The thermodynamic nature of gravity relies on the concept of generalised entropy and heat transfer, as initiated by Bekenstein \cite{Bekenstein} and Hawking \cite{Hawking}. Following this concept, space and time parameters are interpreted as fundamentally thermodynamic variables, in contrast to microscopic parameters. Following \cite{Mandrin1}, it makes sense to formulate a quantum description wich does not depend on space-time at all. As a further consequence, neither Hamiltonian nor Lagrangian formulation of quantum dynamics is required in order to understand the properties of gravity and matter. Rather, dynamics may be considered as a consequence of the thermodynimic behaviour of gravity and matter, and no variational principle needs to be imposed microscopically. In this respect, this approach differs from other background-independent approaches, as no restrictive quantisation procedure is prescribed a priori. In a dynamical microscopic approach, such a restriction would be required, which is not unique. E.g. loop quantum gravity requires quantisation based on the conventional Hamilton-Jacobi formulation of gravity but, also, one may instead choose, say, the precanonical quantisation based on the De Donder-Weyl formulation which has the advantage of being manifestly explicitly covariant while being quantised \cite{Kanatchikov}. 

\paragraph{}
If one decides not to impose any restriction by quantisation, how should quantum measurements be predicted? In MIQG, space-time is a macroscopic parameterisation, and therefore quantum measurements are interpreted as macroscopic measurements on the quantum detectors themselves.

\paragraph{}
In order to describe gravity and matter in a fully general manner, it is convenient and most natural to use the tetrad formalism throughout at the macroscopic level. Thus, the fully general action of gravity and matter shall be rederived accordingly, following a procedure analogous to \cite{Mandrin1}. This then leads to the results presented in \cite{Mandrin2}. Considering fermionic matter as an example, the formal relation between the generalised connection one-form and the fermion spin becomes apparent.

\paragraph{}
This article is organised as follows. The concept of MIQG is first reviewed. To see the complete existence proof and an introduction of the necessary mathematical tools, however, the reader is refered to \cite{Mandrin1}. Then, the generalised pure gravitational action is rederived using the tetrad formalism. Finally, the matter terms are supplemented, and special cases are discussed, especially GR, QFT and fermionic matter.

%%%%%%%%% 2. Overview of the MIQG concept

\section{Overview of the MIQG concept}
\label{sec:2}

Consider portions of ''physical space'' and call each $k$th portion a system $S_k$. Each of the systems $S_k$ is called (macroscopically) separable if 

\noindent 1. one may attribute to it a large enough number $n_k$ of quanta, $n_k \gg \sqrt{n_k}$, 

\noindent 2. it is in thermal equilibrium, and 

\noindent 3. it differs from other systems by at least one macroscopic parameter, e.g. location \cite{Mandrin1}.

\paragraph{}
For asymptotically flat space-times, the ADM-mass as a conserved quantity is an appropriate physical quantity to introduce a first quantum number, namely the ''mass-energy''.

\paragraph{}
As a second step, it is necessary to introduce some sort of ''spatial parameterisation'' which can be done in an arbitrary way. Clearly, the space-time parameterisation applies exclusively to macroscopically separable systems. On the other hand, if the large quantum number condition fails to apply, the system has negligible impact on the macroscopic variables and thus on space-time geometry. 

\paragraph{}
How can quantum measurements be performed in spite of missing quantum space-time dynamics? A quantum measurement involves at least one macroscopic quantity (e.g. location). Thus, the measurement must be performed on a macroscopic system, namely the detector itself. Quantum detectors are in unstable thermal equilibrium, so that tiny changes in quantum number contents in the detector may cause the detector to fire. Firing of a detector is an avalanche process of successive interactions inside the detector, which is an irriversible thermodynamical process \cite{Mandrin1}. Also, wave function collapse is reconsidered from this perspective, as well as the issue concerning unitarity versus increasing entropy. It can also be seen that the memory of one quantum measurement can be discarded by subsequent measurements (e.g. Stern-Gerlach-type experiments).

\paragraph{}
It is shown in \cite{Mandrin1} that QFT directly follows from MIQG in the special case of negligible gravitational field. In particular, the second quantisation has been shown to apply once the macroscopic variation principle has been derived for a given field of many particles. One brings the Langrangian into canonical form (compute the hamiltonian and insert the solution into the Euler-Lagrange equations). Finally, the $k$-mode particle number factor of the hamiltonian may be replaced by the $k$-mode number operator, in terms of particle creation and annihilation operators to simulate the particle emission and detection process ($n_A(k) \rightarrow \hat{a}_k^\dagger \hat{a}_k$). This then yields the operator-valued field. This is explicitly sketched in \cite{Mandrin1} for the example of the Klein-Gordon field.

\paragraph{}
Third, one assumes that the quanta of mass-energy may be distributed arbitrarily within $n_{Sk}$ boxes per system $k$. The mean box filling corresponds to the ''temperature'' $T_k$, while the number $n_{Sk}$ is proportional to the entropy of the system, $S_k = n_{Sk} \ln p$, where $p$ is the (finite) maximum occupation number per box. If the parameterisation of a set $\{S_k\}, k = 1 ... N$, of macroscopically separable systems is chosen to be 1-dimensional, while ignoring interactions, any total ''mass-energy change'' $\delta U_k$ of the $k$th system is simply given by the ''first law of thermodynamics'' ($c=G=k_B=\hbar=1$),

\begin{equation}
\label{eq:first_law1}
\delta U_k = T_k \delta S_k,
\end{equation}

\noindent 
and it is easily shown that, for any (small enough) ''volume cell'' $\Sigma$ in approximate thermal equilibrium, the (quasi-local) first law rerwritten und combined with the second law ($\delta S = 0$) yields

\begin{equation}
\label{eq:var_region1}
\delta S = [T^{-1}(x) \delta \mathcal{U}(x)]\bigg|_{\partial\Sigma} = 0.
\end{equation}

\noindent To obtain Eqn. (\ref{eq:var_region1}), one has to convert the required summation over $k$ into an integral and to integrate by parts. Furthermore, this equation is the variation of the microcanonical action in one dimension, or more accurately, its boundary term. Indeed, the boundary term represents the matter it contains.

The spatial parameterisation may be chosen arbitrarily, i.e. there must be a permutation invariance of the space parameter. One is also free to choose a parameterisation of higher dimensional space, $n > 1$. In this case, it is convient to introduce a topological space $\mathcal{M}$. Using the approximate thermal equilibrium condition over adjacing volumes and the required smoothness condition of temperature (required for hydrodynamics to apply), one easily shows that this topological space is a smooth manyfold, and the formalism thus becomes diffeomorphism-invariant.

\paragraph{}
As is also shown by \cite{Mandrin1}, Lorentzian space-time ($n = 3+1$) optimises the ratio of locally resulting degrees of freedom per imposed degree of freedom (ratio = $16/10)$.

\paragraph{}
In the same way as translation invariance in $n=1$ justifies the assumption of mass-energy conservation a posteriori, global angular momentum conservation may be seen to be generated by rotations (translation of angle-valued coordinates), which leads us to a surface (angular) momentum density contributing to the boundary intregral of the entropy as an additional term in the variation principle, Eqn. (\ref{eq:var_region1}). By the same ananlogy, angular momentum conservation induces the ''angular momentum quantum number''. Finally, diffeomorphism invariance allows us to introduce the usual tetrad formalism as a standard and natural space-time transformation basis, thus leaving the physical laws invariant under these transformations in the most general way.

Considering a thermally small space-time region $\mathcal{M}$ (according to the definition in \cite{Mandrin1}), we are finally lead to a more general form of the second law without interactions, as follows:

\begin{equation}
\label{eq:scaling_no_p}
\delta S\bigg|_\mathcal{T} = \int_\mathcal{T} {\rm d}^3x [ N \delta ({\sqrt{\sigma} \epsilon}) - N^i \delta (\sqrt{\sigma} j_i)],
\end{equation}

\noindent where, without loss of generality, $\mathcal{M}$ has a timelike boundary $\mathcal{T}$ and two spacelike boundaries $\Sigma+$ and $\Sigma-$ normal to $\mathcal{T}$, $N$ and $N^i$ are identified to be the (generalised) lapse and shift, respectively, $\epsilon$ is identified to be the (generalised) surface energy density, $j_i$ corresponds to the (generalised) surface current density, and time integration has been replaced by a locally constant factor because of thermal smallness and using the periodicity of the wick-rotated time. A more detailed derivation may be found in \cite{Mandrin1}. Also, we are using the same notation as in \cite{Mandrin1} for indices, metrics and projectors.

\paragraph{}
Because of the diffeomorphism invariance of space-time, the tetrads also come into play in Eqn. (\ref{eq:scaling_no_p}) (explicit dependence), and hence one has to perform a variation with respect to the tetrads as well. In this way, we obtain one more term and thus the starting point for the computation of the spin-compatible action:

\begin{equation}
\label{eq:scaling_p}
\delta S\bigg|_\mathcal{T} = \int_\mathcal{T} {\rm d}^3x [ N \delta ({\sqrt{\sigma} \epsilon}) - N^i \delta (\sqrt{\sigma} j_i) + N \sqrt{\sigma} s^i_I \delta e^I_i].
\end{equation}

\noindent Comparison with \cite{Mandrin1} and, e.g., \cite{Brown_York_1992} or \cite{Creighton_Mann} suggests to interpret $s^i_I = s^{ij}e^J_j\eta_{IJ}$ as the surface stress one-form, which accounts for a pressure and a deformation contribution (after taking its quadratic form), and supplemented by a ''torsion contribution'' (the remainder). The torsion contribution is important for the computation of the generalised, spin-compatible connection one form, as shown in the next section. Note also that quantities like $\epsilon$ or $j_i$ should not be confused with the corresponding quantities resulting from the ADM-decomposition of GR, as may be found in \cite{Brown_York_1992}, \cite{Brown_York_1993} or \cite{Creighton_Mann}. Rather, $\epsilon$ and $j_i$ are generalisations thereof, and GR has not been used to derive them. 

%%%%%%%%% 3. Derivation of the purely gravitational spin-compatible action

\section{Derivation of the purely gravitational spin-compatible action}
\label{sec:3}

The derivation of the action in tetrad notation follows essentially the same procedure as in metric notation \cite{Mandrin1}.

Because variations $\delta e^I_i$ imply changes of the volume of $\mathcal{M}$, they also affect the volume of neighbouring pieces of space-time. In this respect, $\mathcal{M}$ is not isolated but lies in a ''bath of tetrads''. Thus, we perform a Legendre transformation ($e^I_i \rightarrow N \sqrt{\sigma} s^i_I$) and obtain the ''enthalpy'', so that the surface stress density one-form is varied instead of the tetrads. Then compute the variational derivatives of the integrand $s$ of the action on the boundary $\mathcal{T}$: 

\begin{eqnarray}
\epsilon & = & (\delta s / \delta N) \big|_\mathcal{T}/ \sqrt{\sigma}, \nonumber \\
j_i & = & (\delta s / \delta N^i) \big|_\mathcal{T}/ \sqrt{\sigma}, \nonumber \\
\label{eq:var_derivatives}
s^i_I & = & (\delta s / \delta e^I_i) \big|_\mathcal{T}/ \sqrt{-\gamma}.
\end{eqnarray}

\paragraph{}
As in \cite{Mandrin1}, $\delta s\big|_\mathcal{T}$ is again an exact differential of the function $s\big|_\mathcal{T}$ and there exists a common 1-form ${\tau_\mathcal{T}}^i_I$ of the pure gravity contributions on the boundary which is given by

\begin{equation}
\label{eq:tau_oneform}
\tau^i_I = \tau^{ij} g_{jk} e^k_I = \tau^{ij} \eta_{IJ} e^J_j,
\end{equation}

\noindent where $\tau_{ij}$ is given from \cite{Mandrin1}. Hence, our Legendre transform of Eqn. (\ref{eq:scaling_p}) reduces to

\begin{equation}
\label{eq:var_e}
\delta S\bigg|_{\partial\mathcal{M}} = \sum_{\mathcal{A} = \mathcal{T}, \Sigma} \int_{\mathcal{A}}  {\rm d}^3x \sqrt{|\gamma|} {\tau_\mathcal{A}}^i_I \delta e^I_i,
\end{equation}

\noindent after having included the space-like boundaries $\Sigma$ as well to evaluate the action on the whole boundary $\partial\mathcal{M} = \mathcal{T} \cup \Sigma$. 

\paragraph{}
Performing Legendre transformations in such a way that the total surface stress density 1-form ${\tau_\mathcal{A}}^i_I$ is varied instead of the tetrad now yields:

\begin{equation}
\label{eq:var_tau}
\delta S\bigg|_{\partial\mathcal{M}} = \int_{\partial\mathcal{M}}  {\rm d}^3x \sqrt{\gamma} e^I_i \delta {\tau_\mathcal{A}}^i_I
\end{equation}

This integral over $\partial\mathcal{M}$ may be reexpressed as an integral over $\mathcal{M}$ using the same technique as in \cite{Mandrin1}. We can write ${\tau_\mathcal{A}}^i_I$ as a flow across the boundary, ${\tau_\mathcal{A}}^i_I = {\tau_\mathcal{A}}^{i\gamma}_I n_\gamma$, with $n_\mu$ being the normal unit vector on the boundary. Then we apply Gauss' law,

\begin{equation}
\label{eq:gauss}
\sum_\mathcal{A} \int_{\mathcal{A}} {\rm d}^3x \sqrt{|\gamma|} {\tau_\mathcal{A}}^{i\gamma}_I n_\gamma \delta e^I_i = \int_\mathcal{M} d^4x \sqrt{-g} \nabla_\gamma(e^I_\alpha \delta {\tau_\mathcal{A}}^{\alpha\gamma}_I )
\end{equation}

\noindent By using Leibnitz' rule, the integrand may conveniently be written in terms of variation of the variables conjugate to the tetrads and to the connection 1-form, as follows:

\begin{eqnarray}
\label{eq:dUpsilon}
\Upsilon & = & \nabla_\gamma(e^I_\alpha \delta {\tau_\mathcal{A}}^{\alpha\gamma}_I ) \nonumber \\
& = & e^I_\alpha \nabla_\gamma \delta {\tau_\mathcal{A}}^{\alpha\gamma}_I + (\delta {\tau_\mathcal{A}}^{\alpha\gamma}_I) \nabla_\gamma e^I_\alpha \nonumber \\
& = & e^I_\alpha \eta_{IJ} \nabla_\gamma [g^{\delta\epsilon} e^J_\delta e^K_\epsilon \delta {\tau_\mathcal{A}}^{\alpha\gamma}_K] + \delta {\tau_\mathcal{A}}^{\alpha\gamma}_I \nabla_\gamma e^I_\alpha \nonumber \\
& = & e^I_\alpha e^J_\delta \eta_{IJ} g^{\delta\epsilon} \nabla_\gamma [e^K_\epsilon \delta {\tau_\mathcal{A}}^{\alpha\gamma}_K] + [e^I_\alpha \eta_{IJ} g^{\delta\epsilon} e^K_\epsilon  \delta {\tau_\mathcal{A}}^{\alpha\gamma}_K \eta^{JL} + \delta {\tau_\mathcal{A}}^{\alpha\gamma}_I \eta^{IL}] \ \nabla_\gamma e_{\delta L} \nonumber \\
& = & e^I_\alpha e^J_\delta \delta\Phi^{\alpha\delta}_{IJ} + [e^I_\alpha \eta_{IJ} g^{\delta\epsilon} e^K_\epsilon  \delta {\tau_\mathcal{A}}^{\alpha\gamma}_K \eta^{JL} + \delta {\tau_\mathcal{A}}^{\alpha\gamma}_I \eta^{IL}] g_{\delta\rho} \eta^{MN} e^\rho_M \ \omega_{\gamma NL}  \nonumber \\
\label{eq:var_Upsilon}
& = & e^I_\mu e^J_\nu \delta\Phi^{\mu\nu}_{IJ} + \omega_{\mu IJ} \delta \Omega^{\mu IJ},
\end{eqnarray}

\noindent where 

\begin{eqnarray}
\label{eq:dPhi}
\delta\Phi^{\alpha\delta}_{IJ} & = & \eta_{IJ} g^{\delta\epsilon} \nabla_\gamma [e^K_\epsilon \delta {\tau_\mathcal{A}}^{\alpha\gamma}_K], \\
\label{eq:dOmega}
\delta\Omega^{\mu IJ} & = & (e^J_\alpha \eta^{KI} + e^I_\alpha \eta^{KJ}) \delta {\tau_\mathcal{A}}^{\alpha\mu}_K, \\
\label{eq:omega}
\omega_{\mu IJ}  & = & e^\alpha_I \nabla_\mu e_{\alpha J}.
\end{eqnarray}

The above quantities are the connection one-form $\omega_{\mu IJ}$ and, by comparison with the Palatini formulation of GR, the generalised curvature 2-form $\Phi^{\alpha\delta}_{IJ}$ and the additional variable $\Omega^{\mu IJ}$ conjugate to the connection 1-form. The last term in the last line of Eqn. (\ref{eq:var_Upsilon}) vanishes in torsionless GR, because $\delta\Omega^{\mu IJ}$ is symmetric in the Minkowskian indices and, therefore, only the non-Lorentz part of the connection contributes to the last term.

Eqn. (\ref{eq:var_Upsilon}) is just the boundary contribution to the total gravitational action. The total action including the proper bulk contribution may be obtained by the same procedure as in \cite{Mandrin1} or according to \cite{York}:

\begin{equation}
\label{eq:Total_grav}
S_{\rm total} = \int_\mathcal{M} d^4x \sqrt{-g} [e^I_\mu e^J_\nu \Phi^{\mu\nu}_{IJ} + \omega_{\mu IJ} \Omega^{\mu IJ}]
\end{equation}

\paragraph{}
Eqn. (\ref{eq:Total_grav}) shows that the gravitational action needs to be varied as a function of both the tetrads and the connection 1-form, in precisely the same way as is done in the Palatini action formulation.

%%%%%% 4. Derivation of the spin-compatible action with matter

\section{Derivation of the spin-compatible action with matter}
\label{sec:4}

The matter contribution to the action may be obtained by allowing exchange of quanta between thermally small volumes as discribed in \cite{Mandrin1}. Quanta may only be exchanged by packets of fixed quantum numbers, i.e. as particles.

Consider the number $n_{pA}$ of particles generating a certain interaction of type $A$. The number $n_{pA}$ is a further thermodynamic variable, and its conjugate variable is the chemical potential $\mu_A$. Because, by construction, the ''specific charge'' of a type $A$ interaction particle is a fundamental constant, the number $n_{pA}$ may be replaced by the interaction charge, and hence, by analogy, the current is a conserved quantity, and we have the charge as a new quantum number.

\paragraph{}
For simplicity, in what follows, we consider the approximation of minimal coupling (a generalisation to non-minimal coupling could still be made as a possible extension).

\paragraph{}
Conformly to \cite{Mandrin1}, the chemical potential must be varied, and we thereby denote the general indices $i_1i_2\ldots i_q$ by $I$, $j_1j_2\ldots j_{q+1}$ by $J$, $\gamma_1\gamma_2\ldots\gamma_q$ by $\Gamma$ and $\delta_1\delta_2\ldots\delta_{q+1}$ by $\Delta$:

\begin{equation}
\label{eq:var_matter}
\delta S_{\rm matter}\bigg|_{\partial\mathcal{M}} = \int_\mathcal{T} {\rm d}^3x \Pi^{IA} \delta A_{IA} \sqrt{\gamma} + \int_{\Sigma+} \ldots - \int_{\Sigma-} \ldots,
\end{equation}

\noindent where $A_{IA}$ is the interaction potential and $\Pi^{IA}$ is the conjugate potential.

\paragraph{}
Because of the interactions, there must also be a matter contribution to the stress one-form $s^i_I$ and to the total stress density one-form ${\tau_\mathcal{A}}^i_I$ which, from now on, shall stand for gravity $+$ matter. Therefore, the total variation of the action on the boundaries reads:

\begin{equation}
\label{eq:var_grav_matter}
\delta S\bigg|_{\mathcal{A}} = \int_{\mathcal{A}}  [e^I_i \delta {\tau_\mathcal{A}}^i_I + {\Pi_\mathcal{A}}^{IA} \delta A_{IA}] \sqrt{\gamma} {\rm d}^3x
\end{equation}

After performing one more Legendre transformation, we obtain the variation with respect to ${\Pi_\mathcal{A}}^{IA}$ instead of $A_{IA}$:

\begin{equation}
\label{eq:var_bound}
\delta S\bigg|_{\partial\mathcal{M}} = \int_{\partial\mathcal{M}}  [e^I_i \delta \tau^i_I + \sum_A A_{IA} \delta {\Pi_\mathcal{A}}^{IA}] \sqrt{\gamma} {\rm d}^3x
\end{equation}

Repeating the same steps as in \cite{Mandrin1} for the matter contribution and copying the results of the last section for the gravitational contribution, we finally obtain the total spin-compatible action:

\begin{equation}
\label{eq:action_final}
S_{\rm total} = \int_{\mathcal{M}}  [e^I_\mu e^J_\nu \Phi^{\mu\nu}_{IJ} + \omega_{\mu IJ} \Omega^{\mu IJ} + \sum_A j'^{\Gamma A} A_{\Gamma A} + F'^{\Delta A} F_{\Delta A}] \sqrt{\gamma} {\rm d}^3x,
\end{equation}

\noindent where the (possibly antisymmetrised) covariant derivative of $A^{\Gamma A}$ is the field $F^{\Delta A}$, the divergence of the flow $\Pi^{\Gamma\mu A}$ of $\Pi^{\Gamma A}=\Pi^{\Gamma\mu A}n_\mu$ across the boundary is the generalised current density $j'^{\Gamma A}$ and the flow of $\Pi^{\Gamma A}$ is the generalised field $F'^{\Delta A}$. In the last step, Gauss' Theorem has been applied once again, and the same argument has been applied as usual in order to obtain the bulk variation term, before writing down the total action $S_{\rm total}$.

\paragraph{}
Let us discuss a few special cases of the total action, Eqn. (\ref{eq:action_final}). 

\paragraph{}
If there is no torsion (and the covariant derivative is metric) and the generalised curvature 2-form $\Phi^{\mu\nu}_{IJ}$ is of dimension at most quadratic in the derivative of the tetrads, then $\omega_{\mu IJ}$ is a Lorentz connection, the generalised curvature 2-form is identical to the curvature 2-form of GR and the gravitational action becomes the Palatini action. In this case, one must exclude any fermionic matter with net spin, as will become apparent next.

\paragraph{}
Consider now pure fermionic matter of one single interaction type, {\rm e. g.} pure electrons emitted in a series of many identical processes $W^- \rightarrow \bar{\nu}_e + e^-$. Then, the pure electonic fluid constitutes the following specific total action, if we neglect possible non-linearites of the electron-field (electron-electron interactions) and satisfying the thermally small mass-energy conservation condition (with non-zero invariant mass $m_e$ per particle):

\begin{equation}
\label{eq:action_fermionic}
S_{\rm total} = \int_{\mathcal{M}}  [e^I_\mu e^J_\nu \Phi^{\mu\nu}_{IJ} + \omega_{\mu IJ} \Omega^{\mu IJ} + \bar{\psi_e}\gamma^I e^\mu_I\nabla_\mu \psi_e - m_e\bar{\psi_e}\psi_e] \sqrt{\gamma} {\rm d}^3x.
\end{equation}

\noindent Furthermore, if the gravitational field due to the electrons and other sources are sufficiently weak so that higher order derivatives of the tetrads can be neglected in the gravitational term, the expression for the action may be approximated by the Einstein-Cartan theory. This means that the presence of net electron spin requires non-vanishing torsion. Thus, as one would expect, there is a relationship between the non-Lorentz part of the connection 1-form and electron spin, and this relationship must also hold in the most general case.

%%%%%%%%%%%%% 5. Conclusions

\section{Conclusions}
\label{sec:5}

Using minimum microscopic information as input to the theory, the most general form of quantum gravity description has been derived and formally constructed. Yet the result appears to reproduce the well-established physical laws up to some freedom in the choice of higher order terms in the curvature and particle density expansion of the total action. Moreover, the tetrad formulation appears to be the only one appropriate for describing all the possible types of matter. In full generality, the torsion contribution of the connection 1-form is related to fermionic spin, as one expects.

%%%%%%%%%%%%%%%% Acknoledgements

\section*{Acknowledgements}
I would like to thank to Igor Kanatchikov and to Igor Bulyzhenkov for their support and feedback. I also thank to Philippe Jetzer for hospitality at University of Zurich.
Finally, I would like to thank to Marianne Schwyter and to my father who both encouraged me during the development of my approach to quantum gravity, as well as to several colleagues in Baden (Switzerland).

%%%%%%%%%%%  References

\end{document}